\documentclass[conference]{IEEEtran}
\usepackage[latin9]{inputenc}
\usepackage{amsmath}
\usepackage{amssymb}
\usepackage{graphicx}
\usepackage{esint}

\makeatletter

\usepackage{cite}

\ifCLASSINFOpdf
  \DeclareGraphicsExtensions{.png}
\else
  \DeclareGraphicsExtensions{.eps}
\fi

\usepackage{slashbox}\usepackage{color}

\usepackage{url}

\DeclareMathOperator{\erfc}{erfc}

\makeatother

\begin{document}

\title{Efficient Incremental Relaying}

\author{\IEEEauthorblockN{Muhammad Mehboob Fareed and Mohamed-Slim Alouini
\\
} \IEEEauthorblockA{Computer, Electrical, and Mathematical Science
and Engineering Division\\
 King Abdullah University of Science and Technology (KAUST), \\
 Thuwal, Makkah Province, Kingdom of Saudi Arabia. \\
 \{muhammad.fareed, slim.alouini\}@kaust.edu.sa} %
\thanks{This work is sponsored by King Abdullah University of Science and
Technology (KAUST), Thuwal, Makkah Province, Kingdom of Saudi Arabia.%
} }
\maketitle
\begin{abstract}
We propose a novel relaying scheme which improves the spectral efficiency
of cooperative diversity systems by utilizing limited feedback from
destination. Our scheme capitalizes on the fact that relaying is only
required when direct transmission suffers deep fading. We calculate
the packet error rate for the proposed efficient incremental relaying
scheme with both amplify and forward and decode and forward relaying.
Numerical results are also presented to verify their analytical counterparts. 
\end{abstract}

\section{Introduction}

Cooperative diversity \cite{Laneman2004} is an effective way to take
advantage of other user's antenna resources to improve error rate
performance. User nodes help a source node in communication by forwarding
the received information from the source node to the destination node.
In this way, it provides an independent replica of the directly received
signal. Both direct and relayed signal are combined at destination.
Relaying can be done in different ways. The relaying node can for
instance completely decode and re-encode information before relaying
which is referred as decode-and-forward (DF) relaying. Alternatively
the relaying node can just amplify the received signal before re-transmitting
it towards destination which is termed as amplify-and-forward (AF)
relaying. What follows, we provide a brief review of the different
relaying strategies and we outline the proposed scheme.

In fixed relaying, the relays always forward the received signal after
either amplifying or after decoding and re-encoding \cite{Laneman2004}.
On the other hand, selective relaying \cite{Laneman2004} utilizes
the instantaneous channel information to decide between relay forwarding
and source re-transmission in the second phase. Whenever the source-to-relay
signal-to-noise (SNR) ratio is above a certain threshold, the relay
forwards its received signal towards the destination to achieve the
benefits of cooperative diversity, otherwise the source is instructed
to re-transmit.

In incremental relaying \cite{Laneman2004,Ibrahim2005,Ikki2008},
feedback from the destination about the success or failure of direct
transmission is used. Relay is allowed to forward signal only when
direct transmission fails otherwise source continues with the next
message reducing total time for transmission from two to one time
slot. Incremental relaying results in better spectral efficiency.
Incremental relaying protocols are extensions of incremental redundancy
protocols, or hybrid automatic-repeat-request (ARQ). \textit{Fractional
incremental relaying} (FIR) protocol \cite{Long2009} further improves
spectral efficiency by re-transmitting partial information from relays.
In FIR, the relay divides the received packet into fractions and sends
a fraction if destination is not able to decode the packet correctly
indicated by a negative acknowledgment (NACK) from the destination.
The relay keeps on sending next fraction until it receives an acknowledgment
(ACK) from the destination or a maximum number of the relay transmission
is reached.

Patil investigates in \cite{Patil2012} the use of packet combining
in cooperative diversity scenario. A comprehensive study of bit error
rate (BER) and throughput performance of the scheme in additive white
Gaussian noise (AWGN) and fading channel with and without maximum
ratio combining (MRC) is provided. A cross layer approach is used
to exploit the inherent time diversity in ARQ retransmissions by MRC
of the packet and the retransmission. 

In our proposed scheme, instead of relaying all the packets, the relay
is engaged only for forwarding the packets for which the SNR of the
direct link is very low. A group of packets is sent from the source
in the broadcast phase which is received by both the destination and
the relay. At the end of each broadcast phase, the destination sends
feedback to the relay to indicate the indices of the $M$ packets
with lowest SNR in the group. The relay then forwards only a subset
of packets as instructed and these re-transmitted packets are then
combined with the directly received signal and decoded at the destination
using MRC. Therefor in contrast to classical incremental relaying,
the proposed scheme avoids complete decoding at destination before
feedback is sent. 

In this paper, we also present a performance analysis and derive the
packet error rate (PER) of the proposed scheme for both AF and DF
relaying. The remainder of the paper is organized as follows. Section
\ref{sec:SM} presents the model of the system under consideration
and preliminary results are provided section \ref{sec:PM}. While
section \ref{sec:PA} gives derivation of the PER, the achievable
diversity order and the efficiency of the proposed scheme are discussed
in section \ref{sec:DOE}. Numerical results are given in section
\ref{sec:OWNR} and finally, section \ref{sec:CN} concludes the paper.

\section{System Model \label{sec:SM}}

\begin{figure}
\center \includegraphics[width=3.2in]{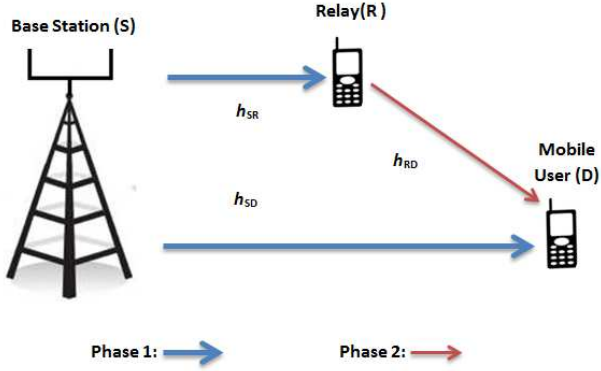} \caption{System model.}

\label{fig:esr2} 
\end{figure}

\begin{figure}
\center\includegraphics[width=3.2in]{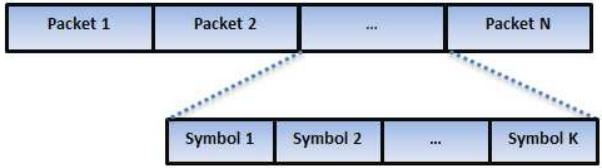} \caption{Frame structure.}

\label{fig:eirframe} 
\end{figure}

We consider a three node system which consists of a source node \textbf{S},
a destination node \textbf{D}, and a relay node \textbf{R}, as shown
in Fig. \ref{fig:esr2}. 
We assume Rayleigh fading channel between transmit and receive nodes
with average SNR for \textbf{S$\rightarrow$R}, \textbf{R$\rightarrow$D},
and \textbf{S$\rightarrow$D} links as $\Gamma_{SR}$, $\Gamma_{RD}$,
and $\Gamma$, respectively. A frame containing $L$ symbols is transmitted
by \textbf{S} towards \textbf{D} which is over heard by \textbf{R}.
Each frame contains $N$ packets each of length $K=L/N$ symbols as
shown in Fig. \ref{fig:eirframe}. 

The channel fading remains constant for the duration of each packet
and may vary from one packet to another. Let $x_{nk}$ be the $k^{th}$
symbol transmitted by \textbf{S} in the $n^{th}$ packet. The signal
received by \textbf{D} is 
\begin{equation}
y_{nk}=\sqrt{E_{s}}h_{SD_{n}}x_{nk}+u_{nk},
\end{equation}
where $h_{n},n=1,2,\cdots,N$ is the channel coefficient during the
$n^{th}$ packet and $u_{nk}$ is the AWGN at the receiver. At the
end of transmission of a frame, the destination sends back the indices
of the $M$ packets for which channel attenuation $|h_{SD_{n}}|$
is worse. Upon reception of index $i$, the relay gets involved in
communication and forwards the $i^{th}$ packet. 

We first consider the AF relaying to re-transmit the signal. In this
case, the relay amplifies the signal 

\begin{equation}
r_{ik}=\sqrt{E_{s}}h_{SR_{n}}x_{ik}+v_{nk}
\end{equation}
 received from \textbf{S} with a fixed gain of 

\begin{equation}
G=\sqrt{\frac{E_{s}}{E_{s}\Gamma_{SR}+N_{0}}}
\end{equation}
 and forwards it to \textbf{D}. The resulting signal received in the
$k^{th}$ time slot of the $m^{th}$ relayed packet at \textbf{D}
is then given by: 
\begin{equation}
z_{mk}^{AF}=h_{RD}Gr_{ik}+w_{mk},
\end{equation}
where $h_{RD}$, $r_{mk}$, and $w_{mk}$ are the channel coefficient
from \textbf{R} to \textbf{D}, the signals received at the relay during
the $m^{th}$ packet, and the AWGN affecting the destination.

We also consider DF relaying, the relays working in DF mode first
decode the received symbols and then re-encode information before
forwarding it to the destination. In order to avoid error propagation
the relay is allowed to participate in the relaying phase only if
the relay is able to correctly decodes the symbol. The signal received
at the destination \textbf{D }in the $k^{th}$ time slot of the $m^{th}$
relayed packet in DF relaying is

\begin{equation}
z_{mk}^{DF}=h_{RD}x_{mk}+u_{mk},
\end{equation}
where $u_{mk}$ the AWGN affecting the destination.

\section{Preliminaries \label{sec:PM}}

In this section, we first provide some results from order statistics.
These results will be used in next section to derive the PER of the
proposed scheme.

Let $\gamma_{i}=|h_{i}|^{2}\frac{E_{s}}{N_{0}}$ be the SNR for the
$i^{th}$ packet, where $h_{i}$ is the channel coefficient, $E_{s}$
is average symbol power, and $N_{0}$ is the noise power. Since Rayleigh
fading channel is assumed, the probability density function (PDF)
of $\gamma_{i}$ is given by: 
\begin{equation}
\begin{matrix}f_{\gamma_{i}}(\gamma)=\frac{1}{\Gamma_{i}}e^{-\frac{\gamma}{\Gamma_{i}}}, & 0<\gamma<\infty,\end{matrix}\label{eq:gi}
\end{equation}
where $\Gamma_{i}=E\left[|h_{i}|^{2}\right]\frac{E_{s}}{N_{0}}=\frac{E_{s}}{N_{0}}=\Gamma$
with average fading power$E\left[|h_{i}|^{2}\right]=1$. We denote
$\gamma_{(i)},i=1,2,\cdots,N$ as the sorted $\gamma_{i}$ such that
$\gamma_{(1)}<\gamma_{(2)}<\cdots<\gamma_{(N)}$. The joint PDF of
$\gamma_{(i)}$'s is known to be given by \cite{Win2001}: 
\begin{equation}
f_{\mathbf{\gamma}}(\gamma_{(1)},\gamma_{(2)},\cdots,\gamma_{(N)})=\frac{N!}{\Gamma^{N}}e^{-\frac{1}{\Gamma}\sum_{i=1}^{N}{\gamma_{(i)}}}.
\end{equation}
The ordered SNRs i.e., $\gamma_{(i)}$'s can be represented by the
sum of independent exponential random variables as \cite{Win2001,Alouini2000}:
\begin{equation}
\gamma_{(i)}=\sum_{m=1}^{i}{\frac{\Gamma}{N-m+1}V_{m}},\label{eq:VB}
\end{equation}
where $V_{m}$ are independent identically distributed normalized
exponential random variables with PDF: 
\begin{equation}
\begin{matrix}f_{V_{m}}(v)=e^{-v}, & 0<v<\infty.\end{matrix}\label{eq:Vm}
\end{equation}
By using the value of $\gamma_{(i)}$ from (\ref{eq:VB}) and by the
taking expectation over $V_{m}$'s we get: 
\begin{eqnarray}
\Phi_{i}(s) & = & E_{V_{m}}\left[e^{-s\sum_{m=1}^{i}{\frac{\Gamma}{N-m+1}V_{m}}}\right]\nonumber \\
 & = & \prod\limits _{m=1}^{i}{\int_{0}^{\infty}{e^{-\frac{s\Gamma}{(N-m+1)}V_{m}}e^{-V_{m}}}dV_{m}}\nonumber \\
 & = & \prod\limits _{m=1}^{i}{\frac{(N-m+1)}{s\Gamma+(N-m+1)}}.\label{eq:perpi}
\end{eqnarray}

\section{Performance Analysis \label{sec:PA}}

Under the assumption, that the fading remains constant over one packet,
the PER for a Binary Phase Shift Keying (BPSK) system with block length
of $K$ symbols on a point-to-point communication link can be calculated
as: 
\begin{equation}
PER(\gamma)=1-\left[1-Q\left(\sqrt{2\gamma}\right)\right]^{K}.\label{eq:perh}
\end{equation}
where $Q\left(.\right)$ is the Gaussian Q-function \cite{Simon2005}.
In order to simplify the above equation we use the approximation of
$\erfc(.)$ given in \cite{Chiani2002} as: 
\begin{equation}
\erfc(x)\approx\frac{1}{2}\left(\frac{1}{3}e^{-x^{2}}+e^{-\frac{4}{3}x^{2}}\right).\label{eq:apxerf}
\end{equation}
Consequently we get 
\begin{equation}
Q(x)=\frac{1}{2}\erfc\left(\frac{x}{\sqrt{2}}\right)\approx\frac{1}{4}\left(\frac{1}{3}e^{-\frac{x^{2}}{2}}+e^{-\frac{4}{3}\frac{x^{2}}{2}}\right).
\end{equation}
We now can calculate the $n^{th}$ power of $Q(x)$ as 
\begin{eqnarray}
Q^{n}(x) & \approx & \frac{1}{4^{n}}\left[\frac{1}{3}e^{-\frac{x^{2}}{2}}+e^{-\frac{4}{3}\frac{x^{2}}{2}}\right]^{n}\nonumber \\
 & = & \sum_{m=0}^{n}C_{n,m}e^{-\frac{A_{n,m}x^{2}}{2}},
\end{eqnarray}
where we define $C_{n,m}={n \choose m}\frac{1}{4^{n}3^{n-m}}$ and
$A_{n,m}=n+\frac{m}{3}$. Expanding Eq. (\ref{eq:perh}) and using
the powers of $Q(x)$, we get the following representation of PER
in terms of sum of exponentials: 
\begin{equation}
PER(\gamma)\approx\sum_{n=1}^{K}D_{K,n}\sum_{m=0}^{n}C_{n,m}e^{-A_{n,m}\gamma},\label{eq:peraprx}
\end{equation}
where $D_{K,n}={K \choose n}(-1)^{n+1}$.

\begin{figure}
\center \includegraphics[clip,width=3.8in]{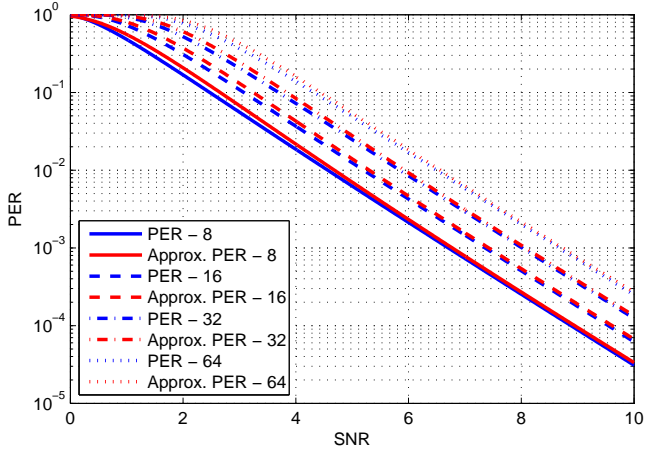} \caption{PER approximation for different values of packet length ($K$).}

\label{fig:apx} 
\end{figure}

Fig. \ref{fig:apx} presents a comparison of the original PER expression
given by Eq. (\ref{eq:perh}) as well as the PER approximation given
by Eq. (\ref{eq:peraprx}). It is obvious that Eq. (\ref{eq:peraprx})
provides better approximation for high SNR and large values of $K$.

The unconditional PER can be calculated by taking expectation over
$\gamma$ yielding 
\begin{equation}
PER\approx\sum_{n=1}^{K}D_{K,n}\sum_{m=0}^{n}C_{n,m}\Phi(A_{n,m}),\label{eq:peraprxu}
\end{equation}
where $\Phi(s)=E_{\gamma}\left[e^{-s\gamma}\right]$ is the Moment
Generation Function (MGF) of the SNR $\gamma$.

\subsection{Packet Error Rate for the Proposed Scheme}

The packet error rate $PER_{T}$ when the relay forwards the $M$-weakest
packets is calculated as 
\begin{equation}
PER_{T}=\frac{1}{N}\sum_{i=1}^{M}PER_{C_{i}}+\frac{1}{N}\sum_{j=M+1}^{N-1}PER_{j},\label{eq:aperk}
\end{equation}
where $PER_{C_{i}}$ is the PER of the $i^{th}$ weakest packet which
is decoded at destination after receiving and combining the direct
and the forwarded signals using MRC and $PER_{i}$ is the packet error
rate for the $i^{th}$ packet. The following two subsections present
now derivations of the PER for AF and DF relaying.

\subsection{Amplify and Forward}

We can calculate the values for the $PER_{C_{i}}$ and $PER_{i}$
for AF relaying using the approximation given by Eq. (\ref{eq:peraprxu}).
If $\gamma_{R}$ and $\gamma_{(i)}$ are the SNR of the relayed packet
and the direct packet with the lowest SNR respectively, $PER_{C_{i}}$
for BPSK can be calculated as: 
\begin{eqnarray}
PER_{C_{i}} & = & PER_{C_{i}}^{AF}\nonumber \\
 & = & E\left[PER(\gamma_{R}+\gamma_{(i)})\right],
\end{eqnarray}
where $E_{X,Y}(.)$ denotes expectation with respect to random variables
$X$ and $Y$. Using Eq. (\ref{eq:peraprxu}), we get: 
\begin{equation}
PER_{C_{i}}^{AF}\approx\sum_{n=1}^{K}D_{K,n}\sum_{m=0}^{n}C_{n,m}\Phi_{C}(A_{n,m})\Phi_{i}(A_{n,m}),\label{eq:pmrcj}
\end{equation}
where $\Phi_{C}(s)$ and $\Phi_{i}(s)$ can be calculated as follows.

By defining $\gamma_{SR}=|h_{SR}|^{2}\frac{E_{s}}{N_{0}}$, $\gamma_{RD}=|h_{RD}|^{2}\frac{E_{s}}{N_{0}}$,
and $c_{1}=1+\frac{E_{s}}{N_{0}}$, for AF relaying with fixed relay
gain, we have \cite{Hasna2004}: 
\begin{equation}
\gamma_{R}=\frac{\gamma_{SR}\gamma_{RD}}{\gamma_{RD}+c_{1}}.\label{eq:gR}
\end{equation}
Consequently, we get: 
\begin{equation}
\Phi_{C}(s)=c_{2}(s)E_{\gamma_{RD}}\left[\frac{{\gamma_{RD}+c_{1}}}{\gamma_{RD}+c_{1}c_{2}(s)}\right],\label{eq:phi}
\end{equation}
with $c_{2}(s)=\frac{1}{1+s\Gamma_{SR}}$. Using \cite[p.366, Eq. (3.384)]{Gradshteyn1994},
we can simplify the above equation similar to \cite[Eq. (35)]{Fareed2008}
as: 
\begin{equation}
\Phi_{C}(s)=c_{2}\left\{ 1+\frac{c_{1}-c_{1}c_{2}}{\Gamma_{RD}}\exp\left(\frac{c_{1}c_{2}}{\Gamma_{RD}}\right)\Gamma\left(0,\frac{c_{1}c_{2}}{\Gamma_{RD}}\right)\right\} ,\label{eq:phic}
\end{equation}
where $\Gamma(.,.)$ is the incomplete Gamma function\cite{Gradshteyn1994}.
Similarly for $PER_{i}$, we can write: 
\begin{equation}
PER_{i}\approx\sum_{n=1}^{K}D_{K,n}\sum_{m=0}^{n}C_{n,m}\Phi_{i}(A_{n,m}),\label{eq:peri}
\end{equation}
where $\Phi_{i}(s)$ is given by Eq. (\ref{eq:perpi}).

\subsection{Decode and Forward}

For a relay system with a decode and forward strategy, we can calculate
$PER_{C_{i}}$ as 
\begin{eqnarray}
PER_{C_{i}} & = & PER_{i}^{DF}\nonumber \\
 & = & PER_{SR}PER_{i}+\left(1-PER_{SR}\right)PER_{C_{i}},\label{eq:pmrcjdf}
\end{eqnarray}
where $PER_{SR}$, $PER_{i}$, and $PER_{C_{j}}$ are the PER at the
relay, the PER at destination when relay is unsuccessful in decoding,
and the PER at destination when relay is successful in decoding, respectively.
The PER for SR link is given by Eq. (\ref{eq:peraprxu}): 
\begin{equation}
PER_{SR}\approx\sum_{n=1}^{K}D_{K,n}\sum_{m=0}^{n}C_{n,m}\Phi_{SR}(A_{n,m}),\label{eq:persr}
\end{equation}
where 
\begin{equation}
\Phi_{SR}(s)=\frac{1}{1+s\Gamma_{SR}}.\label{eq:phisr}
\end{equation}
Similarly, we can calculate $PER_{C_{i}}$ using Eq. (\ref{eq:peraprxu})
\begin{equation}
PER_{C_{i}}\approx\sum_{n=1}^{K}D_{K,n}\sum_{m=0}^{n}C_{n,m}\Phi_{i}(A_{n,m})\Phi_{RD}(A_{n,m}),\label{eq:perci}
\end{equation}
with 
\begin{equation}
\Phi_{RD}(s)=\frac{1}{1+s\Gamma_{RD}}.\label{eq:phird}
\end{equation}

\section{Diversity order and efficiency \label{sec:DOE}}

From Eq. (\ref{eq:perpi}), we observe that for higher values of SNR,
we get:

\begin{eqnarray}
\Phi_{i}(s) & = & \prod\limits _{p=1}^{i}{\frac{(N-p+1)}{s\Gamma+(N-p+1)}}.\nonumber \\
 & \propto & \Gamma^{-i}K_{i},\label{eq:phiprop}
\end{eqnarray}
where $K_{i}$ is a constant of proportionality. Eq. (\ref{eq:phiprop})
eventually leads to 
\begin{eqnarray}
PER_{i} & \approx & \sum_{n=1}^{K}D_{K,n}\sum_{m=0}^{n}C_{n,m}\Phi_{i}(A_{n,m}),\nonumber \\
 & \propto & \sum_{n=1}^{K}D_{K,n}\sum_{m=0}^{n}C_{n,m}\Gamma^{-i}K_{i},\nonumber \\
 & \propto & \Gamma^{-i}\sum_{n=1}^{K}D_{K,n}\sum_{m=0}^{n}C_{n,m}K_{i}.
\end{eqnarray}
Now by putting values of $c_{1}$ and $c_{2}$ in Eq. (\ref{eq:phic})
and using the series representation \cite{Gradshteyn1994} of incomplete
Gamma function similar to \cite[Eq. 35]{Nabar2004}, we can show that
$\Phi_{C}(s)$ decays proportional to $\Gamma$. Hence we can easily
conclude that for AF in the high SNR range: 
\begin{align}
PER_{C_{i}}^{AF} & \propto\Gamma^{-\left(i+1\right)} & i=1,2,\ldots,M\\
PER_{i} & \propto\Gamma^{-i}. & i=M+1,\ldots,N.
\end{align}
Since the least exponent of SNR which corresponds to $i=1$, among
all the terms of $PER_{T}$, is $2$ we can deduce that the diversity
order achieved is $2$ which is also verified in the numerical results
presented in the next section. Similarly we can show that the diversity
order achieved for DF is also 2. 

For efficiency, let us define 

\begin{equation}
\text{FR}=\frac{M}{N}=\frac{\mbox{Number of packets forwarded by the relay}}{\mbox{Total packets}}
\end{equation}
 as the forwarding rate (FR) of the proposed scheme, then we can calculate
the efficiency of the proposed scheme as 
\begin{equation}
\eta=\frac{N}{N+M}=\frac{1}{1+\text{FR}}.
\end{equation}

\section{Numerical Results \label{sec:OWNR}}

\begin{figure}
\center \includegraphics[width=3.8in]{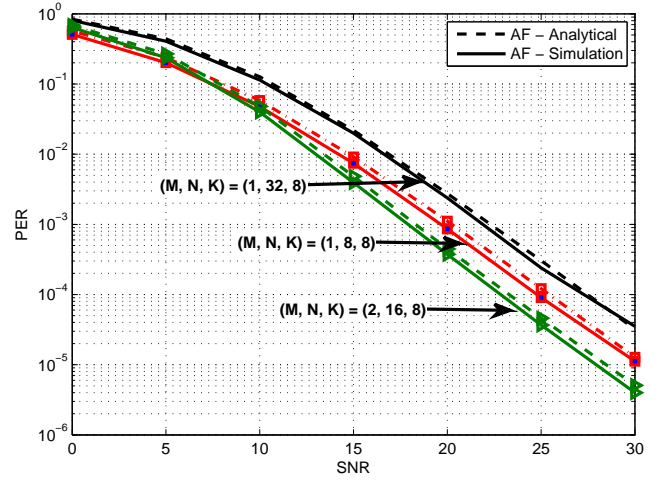} \caption{Comparison of analytical and simulation results for AF with different
values of the ($M$, $K$, $N$) .}

\label{fig:perAF} 
\end{figure}

In this section we present some selected numerical results which were
validated through Monte Carlo simulations of the proposed scheme.
In our simulations, we assume BPSK modulation and a relay which is
placed at an equal distance from both the source and the destination.
Figure \ref{fig:perAF} shows a comparison of analytical PER results
obtained from (\ref{eq:aperk}) and simulated PER for AF. From this
figure, we can observe that the analytical results provide a good
match with those obtained through simulation for different values
of ($M$, $K$, $N$). We observe that Eq. (\ref{eq:apxerf}) results
in very good approximation of the PER for different values of frame,
packet and forwarding rates. It can be further noted that the proposed
scheme performs better with lower value of packet length $K$ and
higher values of $M$ (i.e., number of packets forwarded by the relay).

\begin{figure}
\center \includegraphics[width=3.8in]{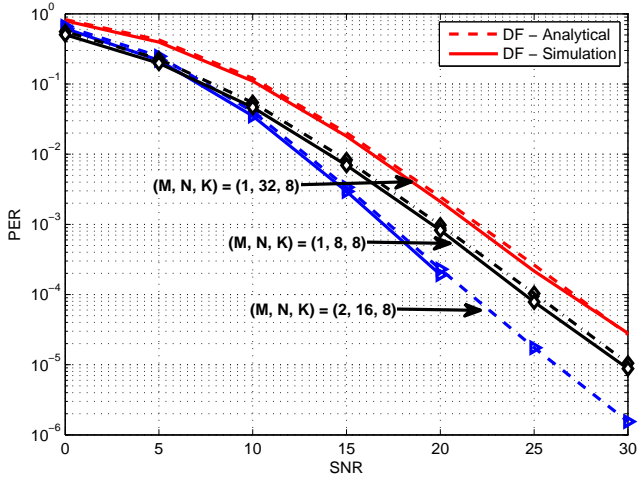} \caption{Comparison of analytical and simulation results for DF with different
values of the ($M$, $K$, $N$) .}

\label{fig:perDF} 
\end{figure}

\begin{figure}
\center \includegraphics[width=3.8in]{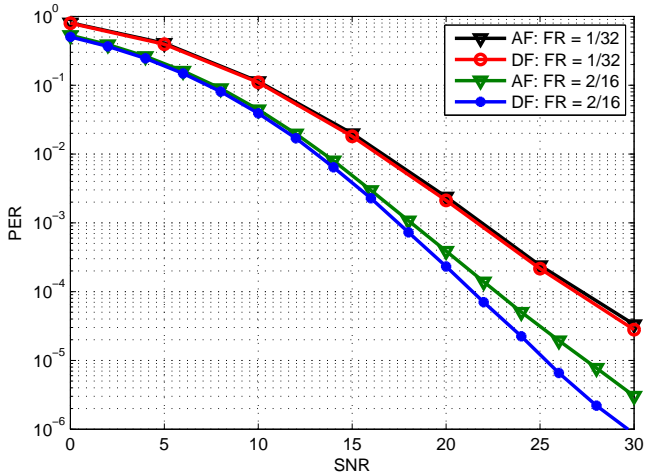} \caption{Comparison of AF and DF with different values of the $M$ .}

\label{fig:perAFDF} 
\end{figure}

Figure \ref{fig:perDF} shows similar comparison of analytical PER
and simulated PER for DF. Again note that the analytical results match
with those obtained through simulation for DF. 
In Fig. \ref{fig:perAFDF} we present a comparison of AF and DF relaying,
we can observe that for $M=1$, DF performance is slightly better
than AF, however DF outperforms AF for $M>1$. 

\begin{figure}
\center \includegraphics[clip,width=3.8in]{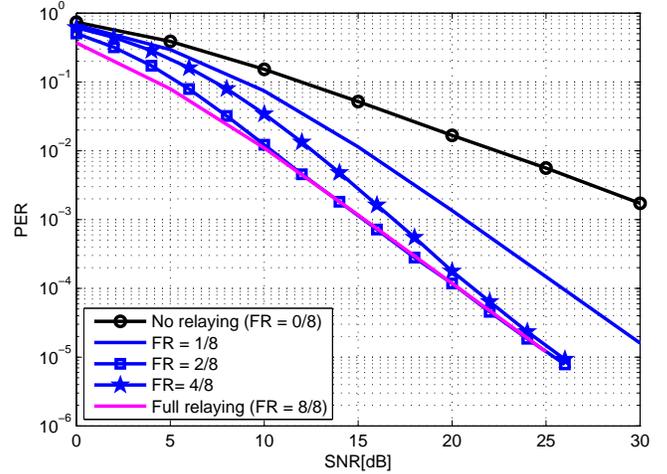} \caption{PER for different forwarding rates $\text{FR}$ with DF relaying.}

\label{fig:esr3} 
\end{figure}

Fig. \ref{fig:esr3} presents some numerical results of the proposed
scheme with $K=16$ and different forwarding rates $\text{FR}$ to
demonstrate the ability of the proposed scheme to achieve better performance
with limited involvement of the relay. 
It can be observed that with a forwarding rate of $\text{FR}=1/8$,
we can get more than $6$ dB improvement at a target PER of $10^{-2}$
in comparison to no relaying (i.e., $M=0$). Note also that the PER
performance at $\text{FR}=4/8$ forwarding rate approaches that of
conventional schemes in which packets are forwarded unconditionally
(i.e., $M=N$) in the medium and high SNR range.

\section{Conclusion \label{sec:CN}}

An efficient relaying scheme was proposed for incremental relaying
which achieves better throughput with the help of minimum feedback
from the destination. Performance analysis based on the packet error
rate is presented for the proposed scheme. The results have shown
that performance of the proposed scheme with partial relay involvement
can approach that of the classical scheme which always uses relay
for forwarding all the information.

 \bibliographystyle{IEEEtran}
\bibliography{IEEEabrv,work_bib}


\end{document}